\documentclass[fleqn,10pt]{wlscirep}
\usepackage{setspace}
\usepackage{url}
\title{Adhesion of the electrodes on diamond device surfaces}
\author[1,*]{Tom Ichibha}
\author[2,3,4]{Kenta Hongo}
\author[5]{I. Motochi}
\author[6]{N. W. Makau}
\author[7]{G. O. Amolo}
\author[2]{Ryo Maezono}
\affil[1]{School of Materials Science, JAIST, Nomi, Ishikawa, Japan}
\affil[2]{School of Information Science, JAIST, Nomi, Ishikawa, Japan}
\affil[3]{National Institute of Materials Science, Tsukuba, Ibaraki , Japan}
\affil[4]{PRESTO, JST, Kawaguchi, Saitama, Japan}
\affil[5]{Department of Mathematics and Physical Sciences, Maasai Mara University, Narok, Kenya}
\affil[6]{Computational Materials Science Group, Department of Physics, University of Eldoret, Eldoret, Kenya}
\affil[7]{Department of Physics and Space Science, The Technical University of Kenya, Nairobi, Kenya}
\affil[*]{ichibha@icloud.com}

\newpage
\begin{document}
\maketitle
\section*{ABSTRACT}
\vspace{3mm}
{\small
\begin{spacing}{1.0}
  Appropriate candidates of the metallic sheet 
used for the electrodes of diamond semiconductor 
are investigated using computational approaches 
based on density functional theory (DFT).
For twenty kinds of metallic elements $x$,
we modeled a diamond-metal interface and evaluated its work of separation,
$W_\mathrm{sep}(x)$, as a possible measure of anti-peeling strength.
The appropriateness of the Ohmic contact was inferred from DOS (density of states) analysis
of diamond-metal interface by looking at whether an in-gap (isolated/localized) peak disappears
as well as a sufficient amount of DOS value exists around the Fermi level.
Our DFT simulation confirmed that a typical electrode, Au, is not adhesive enough for power devices
[$W_\mathrm{sep}(\mathrm{Au}) = 0.80$ J/m$^2$], though showing the Ohmic contact.
In contrast, some transition metals were found to possess Ohmic features
with much stronger adhesion than Au~
[{\it e.g.},~$W_\mathrm{sep}(\mathrm{Cr}/\mathrm{Ti}) = 6.02/4.03$ J/m$^2$].
\end{spacing}
}
\section{Introduction}
\label{sec.intro}
Diamond is a promising candidate for the next-generation
power devices, 
possessing a wider band gap by a factor of five and a
higher thermal conductivity
by a factor of more than ten, \cite{2007MAE}
compared with Si.\cite{2006WILL,2010MAE}
These properties are key for the device robustness
in its anti-voltage and anti-thermal strengths
which is required especially in power semi-conductor devices.
Its higher electron mobility, 1.3 times or more
faster than Si,
makes the device a very good candidate for use in signal processing
with higher frequencies.

\vspace{2mm}
Chemical vapor deposition (CVD) has enabled industrial synthesis of diamond
for devices,\cite{1982MAT,1983KAM}
leading to the recent feasibility to obtain larger single crystals
with higher purity.\cite{2010MOK}
Doping techniques to establish the junctions have also
been studied intensively both from theory and experiments.\cite{2010MOK}
The $p$-type doping has been well developed and
Metal-insulator semiconductor field-effect transistor (MISFET)
has been realized
on the films by homo epitaxial growth.\cite{1999UME}
Diamond junction FET have indeed been
confirmed to work very well.\cite{2012IWA}
In contrast, the $n$-type doping has been difficult to realize, 
but its synthesis has recently been reported.\cite{2005KAT}
The feasibility for device fabrication is now quite real thereby leaving
the matter of the realization of $n+$ layer in the doping challenges.

\vspace{2mm}
Another important issue is the electrode fabrication,
namely the Ohmic contact of the metallic layer on the diamond surface.
There are several experimental attempts that have been made,
mainly using Au because of its low contact resistance.
\cite{1995HEW,2000LOO,1997GLU}
For $p$-type doped diamonds the electrode fabrication
has been well established as reported.\cite{1988MOA}
Although diamond surfaces are naturally clean,
{\it i.e.}, not terminated with either hydrogen or oxygen,
the possible varieties of the surface terminations
and the dependence of the electrode properties on
the termination elements are of great interest to researchers.
\cite{2000LOO,1997GLU,1994AOK,2010CHE,1991MOR,1996KAW}
For instance, it is experimentally reported that 
diamond surface-channel FET's on hydrogen-terminated surface
would be promising candidates for some power applications.\cite{1997GLU}
Other metallic elements especially Al, Au/Ti, and Al/Ti,
have systematically been investigated to establish
if they form better electrodes on the hydrogen-terminated $p$-type
diamonds fabricated by CVD.\cite{2000LOO}
Besides typical electrode metals such as Au and Pd, 
both Ti and Mo have been used for electrode fabrications and
their specific contact resistance has been measured
experimentally\cite{2000LOO}
using circular transfer length method (c-TLM).\cite{1995HEW}

\vspace{2mm}
In order to realize diamond power devices,
it is quite important to establish how to fabricate
such electrodes that
possess not only low contact resistance\cite{1995HEW},
but also high adhesion to the surface.
This is essential for establishing anti-peeling strength
under high voltage and temperature that may occur
in power devices.\cite{2006WILL}

\vspace{2mm}
A number of theoretical works have treated the adhesion between
diamond surface and several metals, \cite{1988PICKETT,1989LAMBRECHT,1991ERWIN,2001WANG,2003QI,2004QI,2005JIA,2010GUO,2012MOTOCHI,2012TIWARI,2014MONACHON}
but their exploration
space were limited as shown here and hence there still exist
the possibilities of discovering more appropriate electrode
metals for diamond surfaces.
Pickett and Erwin \cite{1988PICKETT,1991ERWIN} 
were among the first to investigate metal/diamond interfaces
for electronic device applications 
using first-principles local-density functional approaches,
followed by the pioneering work on BN/diamond interfaces by
Lambrecht and Segall.\cite{1989LAMBRECHT}
They modeled nickel/diamond interfaces on both the ($001$)
and ($111$) ideal surfaces and 
computed their Schottky barrier heights and interface energies. 
It was found that the tetrahedral arrangement leads to an Ohmic interface 
with the interface energy of 0.97 eV per carbon atom, 
suggesting that the interface geometry plays a crucial
role in its surface electronic structure. 
Afterward, the Ohmic properties were verified by an
experimental work~\cite{1994WEIDE}
to support the above theory-driven prediction.

\vspace{2mm}
Recent works~\cite{2010GUO,2012TIWARI,2012MOTOCHI,2014MONACHON}
examined material structure model and studied several metals.
Guo {\it et al.} \cite{2010GUO} treated 3 metals(Al, Cu, Ti)
on clean diamond ($111$)-($1\times1$) surface.
Monachon {\it et al.} \cite{2014MONACHON} treated 2 metals (Cu, Ni) 
on clean and hydrogen terminated diamond ($111$)-($1\times1$) surfaces.
These two works evaluated the adhesion by the work of separation ($W_{\mathrm{sep}}$),
and they treated the model of metallic electrodes as multi-layer:
The work of separation represents the adhesion between metal and diamond bulks.

\vspace{2mm}
On the other hand, Motochi {\it et al.} \cite{2012MOTOCHI} modeled
the interfaces as periodic slabs comprised of monolayer metallic electrodes 
and ten-layer diamond surfaces with/without monolayer atomic terminations.
They applied density functional theory (DFT) approaches to evaluate adsorption
energy, $E_{\mathrm{ads}}$ and density of states (DOS) for their target systems.
They concluded that tantalum and vanadium were the best metallic electrodes
  because they showed highest $E_{\mathrm{ads}}$
  with surface metallic properties,
  {\it i.e.} the localized surface electronic state,  
  after the adhesion.
%

\vspace{2mm}
In this study we revisit the previous work by
Motochi {\it et al.} \cite{2012MOTOCHI}
because of the following points. 
Although their work was a challenge and gave a great insight
to the possibility of the carbide forming metals 
to form better electrodes,
their computations and metals considered were limited.
In contrast, the present study has systematically explored
a broad range of metals for more desirable diamond electrodes.
We considered twenty types of elements: 
Mg, Al, Ti, V, Cr, Ni, Cu, Zn, Zr, Nb,
Mo, Pd, Ag, In, Hf, Ta, W, Pt, Au, and Pb.
Here we adopted the same metal/diamond structure model as
that by Motochi {\it et al.}, \cite{2012MOTOCHI}
but paid more careful attention to the evaluation of adhesion
for which we employed $W_\mathrm{sep}$ instead of $E_\mathrm{ads}$.
This is because $E_\mathrm{ads}$ is inappropriate for our purpose as discussed later.
Our more extensive exploration with great care has led us to a different conclusion
from that of Motochi {\it et al.} \cite{2012MOTOCHI} in terms of the best choice for metallic electrode.

\vspace{2mm}
In this section we have reviewed previous studies
on the diamond surfaces and made a brief explanation of our findings.
The following sections are organized as follows:~
\S\ref{sec.det.spec} presents the surface models and methodology.
In \S\ref{sec.results}, the results of the work are presented 
and finally the conclusions drawn from
the findings are given in \S\ref{sec.conc}.
 
\section{Model and Methodology}
\label{sec.det.spec}
\begin{figure}
 \begin{center}
  \includegraphics[width=0.5\hsize]{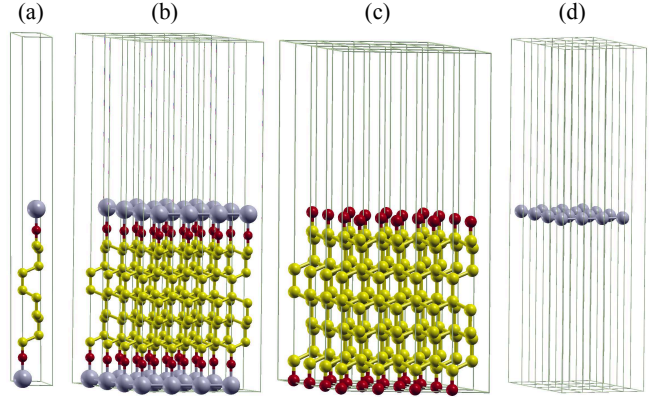}
  \caption{Interface models: (a) unit cell of slab,
    (b) $4 \times 3 \times 1$ supercells of slab,
    (c) parent surface (with terminations), and (d) metallic electrode. 
    The large, small (dark), and small (light color) balls correspond to
    the metal, terminating atoms (hydrogen or oxygen),
    and carbon atoms, respectively.}
  \label{fig.model}
  \end{center}
\end{figure}

Fig.~\ref{fig.model} shows our slab model of
interfaces between metallic electrodes and diamond
($111$)-($1\times1$) surfaces
with/without terminations (H and O), which is the same as 
that in the previous work by Motochi {\it et al.}.\cite{2012MOTOCHI}
We considered the 20 metals and three terminations
and hence a total of 60 interface systems in order to
explore desirable electrodes
for power device applications.
We have performed DFT calculations with
carefully chosen computational conditions (shown later) 
and optimized the geometries using
{\sc Quantum ESPRESSO}~\cite{quantumespresso}
with PBE-GGA~\cite{PhysRevLett.78.1396} exchange-correlation XC functional
and ultrasoft pseudo potentials~\cite{PhysRevB.62.4383} available therein.
  Most of the preceding works have taken
  $W_\mathrm{sep}$ as the measure of adhesion strength, 
  \cite{2001WANG,2003QI,2004QI,2010GUO,2014MONACHON}
  which excludes the energy gain
  due to the cohesion of the metallic sheet itself
  unlike $W_\mathrm{ads}$.\cite{2012MOTOCHI}
  Thus, we also adopted $W_\mathrm{sep}$ to
  evaluate the adhesion strength.
%
We have verified whether or not our model can accurately simulate the interfaces
in terms of partial density of states (pDOS) at each of the diamond layers,
which was not investigated in the previous study.\cite{2012MOTOCHI}

\vspace{2mm}
We define the work of separation:
\begin{equation}
  {W_{{\rm{sep}}}}\left( x \right) = \frac{1}{{2\Omega }}\left( {{E_{{\rm{parent\_surface}}}} + 2 \cdot {E_{{\rm{sheet}}}}\left( x \right) - {E_{{\rm{slab}}}}\left( x \right)} \right).
  \label{w_sep}
\end{equation}
Here, $\Omega$ is the surface area of unit cell,
(i) $E_{\mathrm{parent\_surface}}$ is the energy per unit cell
of the diamond slab with/without termination (parent surface),
(ii) $E_{\mathrm{slab}}(x)$ is the energy of the whole interface system (slab)
with the sheets of metal $x$, and
(iii) $E_{\mathrm{sheet}}(x)$ is the energy of the sheet of metal $x$ (electrode).
The factor $1/2$ in eq.~\eqref{w_sep} comes from
the fact that our slab model has
interfaces on both sides.

%

\vspace{2mm}
Within the framework of the DFT approaches,
all the three energies (i), (ii), and (iii) were computed
under the conditions that $E_{\mathrm {ecut}}^{\mathrm {wfc}} = 90$ Ry
for the cutoff energy of the orbital function expansion and
$E_{\mathrm {ecut}}^{\mathrm {rho}} = 520$ Ry to compensate charges
in the ultrasoft pseudo-potential evaluation.
The $k$-mesh size (discretization of Brillouin zone) of $14\times14\times1$ 
was used for all the three energies.
The Marzari-Vanderbilt smearing scheme~\cite{PhysRevLett.82.3296}
with $\delta E = 0.02$ Ry was applied to all the systems.
The above computational specifications were the best   
choices for all the systems,
but they were also very carefully chosen such that all the
evaluations of interface energy
lie within the chemical accuracy of $\sim 2$ mRy/unit cell.
Hereafter we describe how to model 
the metal-(terminating atom)-diamond interface system (slab) 
and its subsystems (parent surface and metal) in order to evaluate 
the corresponding three energies in more detail.

\vspace{2mm}
First, we extracted a ten-layer diamond slab 
with an ideal ($111$)-($1\times1$) surface from the bulk structure.
As confirmed later in \S~\ref{subsec:layer}, 
the number of layers is sufficienctly large to capture 
the change in the electronic structure from 
surface to bulk inside. 
To construct the diamond surface, we chose a vacuum phase with
dimension of 9.2 \AA 
(\ref{fig.model}) subject to the periodic boundary condition, 
which means that the upper and lower five layers are identical.
In order to take into account surface reconstructions,
we optimized both the atomic positions and lattice parameters
(i.e., unit cell size) simultaneously 
within the PBE-GGA method under the above condition.
Our resulting optimized geometry reasonably agrees with
the experiment~\cite{PhysRevB.66.201401} as well as that by Motochi {\it et al.}.
\cite{2012MOTOCHI}
The number of the layers (Fig.~\ref{fig.model}) has been found
to be large enough to simulate the ($111$)-($1\times1$) surfaces
because the geometry at the fifth and sixth layers
is almost the same as that of the bulk.
The optimized geometry was used to evaluate $E_{\rm parent\_surface}$.
Erwin and Pickett~\cite{steven} pointed out that
the dangling bonds of the ($111$)-($1\times1$) surface are located
at on-top sites.
In the case of H- and O-terminations, therefore,
we placed the terminating atoms at an on-top site of the above optimized
($111$)-($1\times1$) surface.
Starting with initial atomic configurations separated by their
covalent radii,\cite{2008COR}
we fixed the lattice parameters and reoptimized all the atomic positions
and found that both the hydrogen and oxygen terminating atoms
stay at the on-top site with only a change in the vertical distance
between the surface carbon atom and the terminating atoms.

\vspace{2mm}
Secondly, since the on-top site may be thought of as being the
most preferable to electrode adhesion,
we constructed the slab model by putting metallic monolayers
at the on-top site of the optimized ($111$)-($1\times1$) surface.
Starting with initial atomic configurations separated by
their covalent radii,
we fixed the lattice parameters and reoptimized all the atomic
positions to get $E_{\rm slab}(x)$.
Note that our unit cell contains only one metallic atom on the surface. 
This means that we ignore lattice mismatches at interfaces,
but as discussed later, they were found to be negligible.
Similar to the no-termination case, we placed metallic
monolayers on the H- or O-terminating diamond surfaces
and reoptimized the geometries. 

\vspace{2mm}
Finally, we considered the metallic monolayers
({\it i.e.}, two dimensional metallic sheets) 
without optimizing their geometries,
which was used to evaluate $E_{\mathrm{sheet}}(x)$.
The errors due to omitting the optimizations 
are found to be negligible as discussed later 
in \S~\ref{subsec:work}.
Since all the slab systems are non-magnetic,
we treated all the 2D sheets as being
paramagnetic, irrespective of their actual
magnetic states.\cite{2004HON}

\vspace{2mm}
The DOS analysis gives a useful insight into the
Ohmic contact property, 
especially by investigating whether or not the in-gap
peak disappears \cite{2012MOTOCHI} as summarized in 
Fig.~\ref{fig.dos}.
The peak corresponds to the localized 
surface state forming the Schottky contact 
rather than the Ohmic contact. 
The appropriateness for the Ohmic contact 
is therefore
inferred from DOS analysis of metal/diamond interface by checking
if the system possesses 
the larger DOS at $E_F$ (the larger availability 
for valence electrons) and non-peaky shape 
of DOS (non-localized property of electrons). 
In addition, pDOS tells one which angular component contributes
mostly to the carrier in the vicinity of the Fermi level.
We also evaluated the difference between the DOS of diamond slab
and that of the system with electrodes attached in order to
understand how charge transfer occurs.

\section{Results and discussion}
\label{sec.results}
\subsection{Number of diamond layers}
\label{subsec:layer}
\begin{figure}
  \begin{center}
  \includegraphics[width=0.5\hsize]{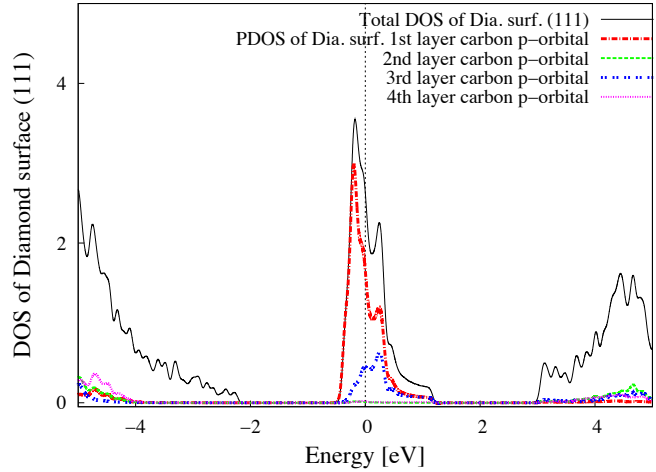}
  \caption{
    Partial contributions to DOS from each layer of
    the diamond slab. Only the first three layers contribute mainly
    to the surface tate appearing between the bulk gap.
    Energy as the horizontal axis is set with the Fermi energy at 0 eV.
  }
  \label{fig.diaallpdos}
  \end{center}
\end{figure}

We have checked if the number of diamond layers
is large enough to simulate
the interfaces by looking into pDOS contributions
at each diamond layer to the total DOS,
as shown in Fig.~\ref{fig.diaallpdos}.
It is found that 2$p$ orbitals from the first three layers
mainly contribute to the surface state appearing within the bulk gap.
This corresponds to the fact that any surface reconstruction
occurs only within the first two layers.
This is consistent with previous studies by Pickett and
Erwin. \cite{1989PICKETT,1991ERWIN}
We note in Fig.~\ref{fig.diaallpdos} that the width of
the in-gap peak, $\sim$ 2 eV, is a bit wider than that
usually expected for the surface state.\cite{himpsel81}

\setlength\intextsep{2pt}
\setlength\textfloatsep{20pt}

\subsection{Density of states}
\label{subsec:dos}
\begin{figure*}
  \begin{center}
  \includegraphics[height=\vsize]{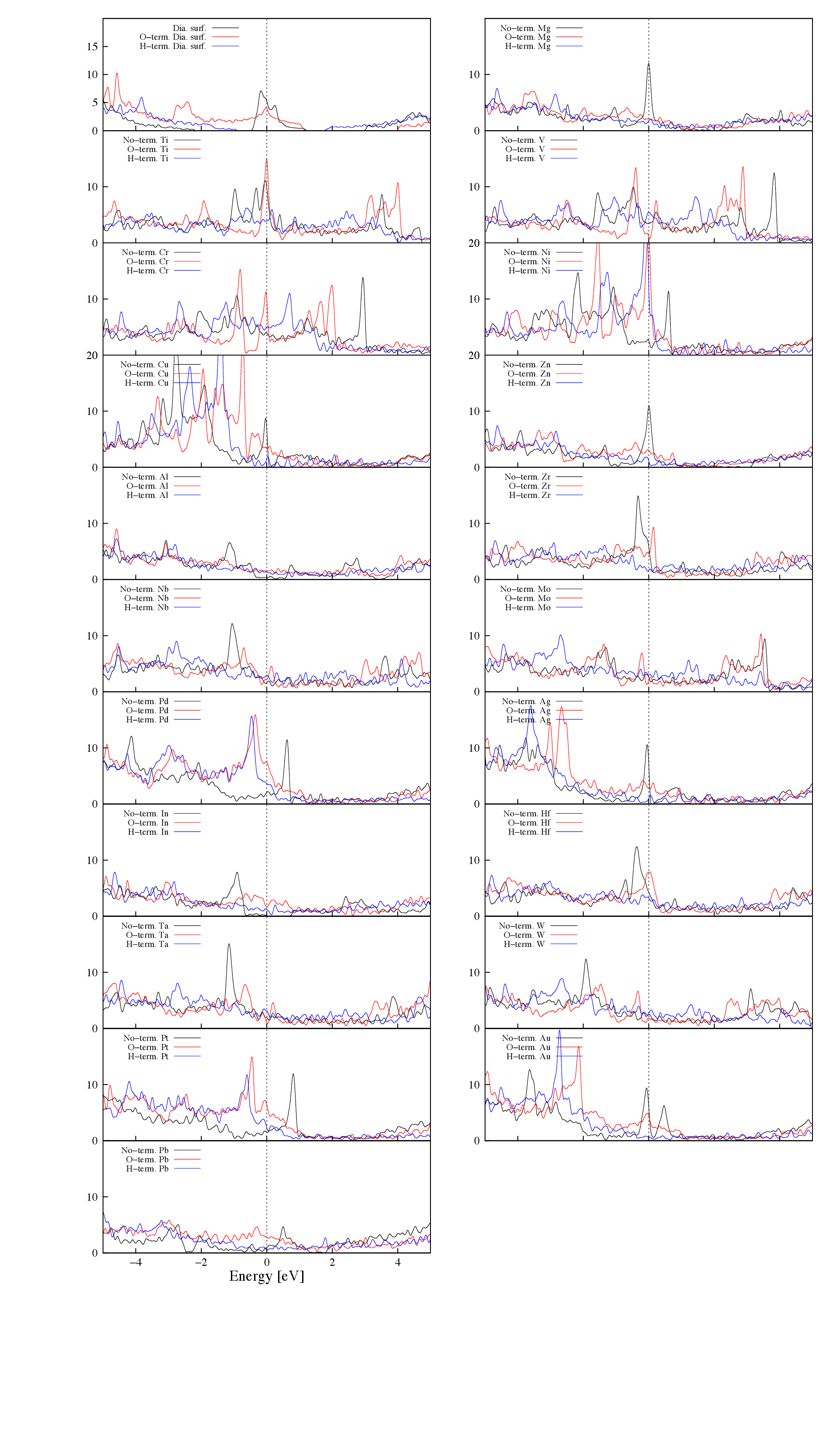}
  \caption{
    DOS of the parent surface and the metal sheets with
    Hydrogen, Oxygen, and no terminations.
    Blue, red, and black lines  correspond
    to Hydrogen,
    Oxygen, and no terminations respectively.
    The vertical dotted line represents Fermi energy.
  }
  \label{fig.dos}
  \end{center}
\end{figure*}

Fig.~\ref{fig.dos} draws the comparison of DOS 
among the different kinds
of metallic sheets to be examined. 
Each panel includes the comparison among 
the kind of terminations (with H, O, and w/o termination). 
The evaluation for the better electrode performances 
is made, as explained in \S\ref{sec.intro}, based on 
(1) whether the DOS fills the gap seen for pristine 
diamond surface in Fig.~\ref{fig.diaallpdos}, and 
(2) how large the DOS is around the Fermi level. 
'Ti sheet w/o termination' therefore achieves 
the best performance in this sense, followed by 
'Cr with H-termination', 'Cr w/o termination', 
'V with H-termination', 'V w/o termination', 
'Cr with O-termination', and 'Ti with O-termination' 
in this order. 
In contrast to that, the familiar electrode materials, 
Au, Pb and Pt, are found to be not necessarily 
the best choice to form the Ohmic contact: 
For Au and Pt, the DOS filling on the gap is 
relatively small. 
Pb achieves acceptable filling but the DOS 
around $E_F$ is relatively small. 
Though the largest DOS around $E_F$ is achieved by 
Ni sheet in a peaky shape, it cannot fill the gap well, 
corresponding to the formation of the localized surface state. 

\vspace{2mm}
Peaky shape of DOS near to $E_F$ in Fig.~\ref{fig.dos}, 
appearing for Ti, Cu, Pd, In, Ta, Pt,
Mg,V, Ni, Zn, Zr, Ag, Hf, and Au, 
would be attributed to the in-gap state
for the diamond slab, but this could be also interpreted as
MIGS (Metal Induced Gap States)~\cite{1965HEI,1984TER}:~
The metallic wavefunction penetrates towards diamond as an evanescent wave
at the metal/diamond interface, and the wave builts the interface states
called MIGS.\cite{1965HEI,1984TER}
\subsection{Bonding natures and adhesion lengths}
\label{subsec:bonding_nature}
\begin{figure}
  \begin{center}
  \includegraphics[width=0.5\hsize]{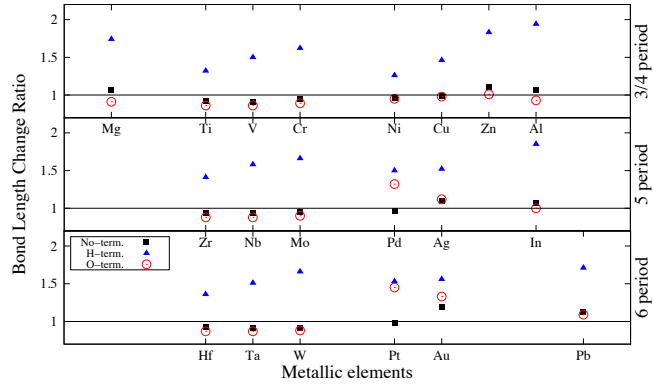}
    \caption{
      Deviations of the adhesion lengths from the initial
      values for the geometrical optimizations. The sum of
      the initial covalent radii is shown to be unity. Open rectangles
      (closed triangles) correspond to the lengths between
      the termination elements and metallic elements (surface carbon atoms).
        No-term. stands for no termination while H-term. and O-term.
        stand for hydrogen- and oxygen- terminated surfaces respectively.
    }
    \label{fig.bondlengthratio}
    \end{center}
\end{figure}

  Fig.~\ref{fig.bondlengthratio} shows the change in the bonding lengths during the lattice relaxation 
  starting off the initial value taken as the sum of covalent radii of 
  constituent atoms of the interface. 
  There is a clear contrast to get elongation or contraction of 
  the adhesion lengths
  after the change between the initial and optimized bond lengths
  (atomic positions were optimized from their initial 
  ones separated by the covalent radii of atoms 
  constituting the interfaces, as explained in \S\ref{sec.det.spec}).
  The contrast might be attributed to the different bonding nature, 
  such as ionic or covalent. 
  In the case of no-termination, for instance, Al (Ti) gives 
  elongation (contraction), being consistent with 
  the covalent (ionic) nature of the elements. 
  Significant elongations for the hydrogen terminations compared 
  with the oxygen terminations and no-terminations 
  might also be attributed to the fact that the
  diamond surface has only one dangling bond that hybridizes with the hydrogen 
  $1s$ orbital and then there is no room for additional bonding with the 
  electrode metals.
In the case of O-termination,
$d_{\mathrm{adh}}$ got contracted [elongated] 
for transition metals (Ti, Ta, V) [noble ones (Au, Pd)].
This contrast can be attributed to whether the d/f orbitals
of the valence shell are closed (for noble metals) or not.
\footnote{
    For Au, the partially occupied 6s orbital can contribute to
    the bonding with oxygen atoms.
    However it may be weak because s orbital has a spherical shape
    and may have less overlap with p orbital of oxygen than
    d/f orbitals.
}
The trend in the magnitude of the contraction can be 
explained to some extent in terms of the electronegativity: 
the more contractions by the elements with the smaller numbering  
of groups ({\it i.e.}, left hand side in the periodic table) 
would be accounted for by the smaller electronegativities. 
The difference of the negativities between the metallic elements 
and the top-most atom on the surface (carbon or oxygen) gets 
larger when the negativity of the metallic element itself gets 
smaller. 
The larger difference leads to the ionic bonding nature, 
and hence to the more contraction. 
The scenario is consistent with the fact that 
we get more contractions by O-termination than no-termination 
because oxygen has the larger electronegativity and hence gives a 
larger difference of electronegativity.

\subsection{Work of separation}
\label{subsec:work}
\begin{figure}
  \begin{center}
    \includegraphics[width=0.5\hsize]{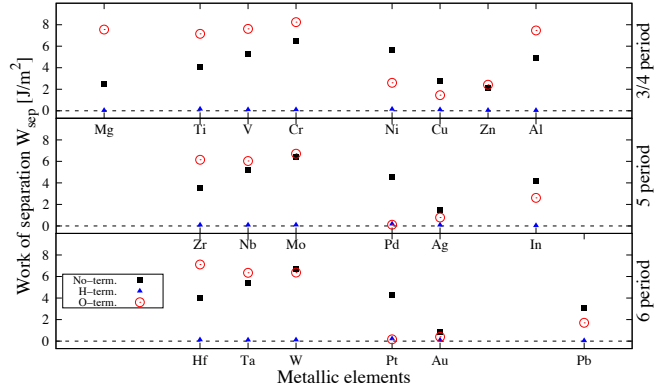}
    \caption{Work of separation
      $W_{\mathrm{sep}}$ obtained from our PBE-GGA simulations
      with the ultrasoft pseudopotentials.
        No-term. stands for no-termination while H-term. and O-term.
        stand for hydrogen- and oxygen- terminated surfaces respectively.
    }
    \label{fig.w_sep}
    \end{center}
\end{figure}

\begin{figure}
  \begin{center}
  \includegraphics[width=0.5\hsize]{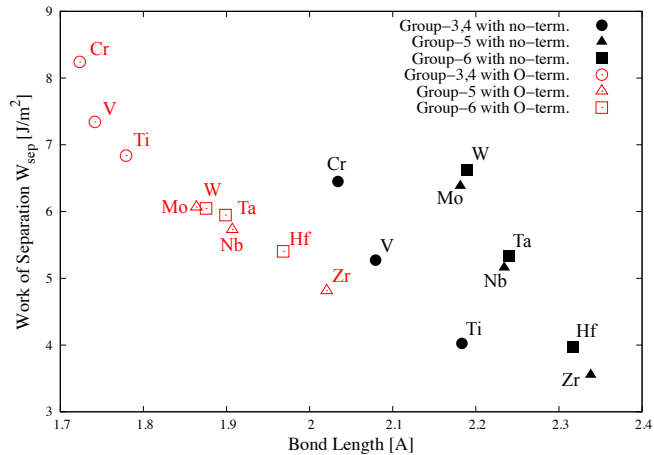}
  \caption{
    Relationship between adhesion length and work of separation $W_{\mathrm{sep}}$.
      No-term. stands for no-termination while O-term. stands for oxygen-terminated surface.
  }
  \label{fig.length_wsep}
  \end{center}
\end{figure}

Fig.~\ref{fig.w_sep} highlights the dependence of
$W_{\mathrm{sep}}(x)$ on the metallic elements $x$
and the terminations (no-termination, O- and H- terminations), 
evaluated from our PBE-GGA simulations with the ultrasoft
pseudopotentials.
It is found that overall trends in 
$W_{\mathrm{sep}}(x)$ strongly depends on the termination 
elements, reflecting their surface bonding natures. 
In the case of the no-termination, as is expected, 
the Au electrode has the weakest adhesion. 
Surprisingly, the Pd is found to have a slightly
stronger adhesion than the Ti that forms a carbide. 
The group-6 elements (Cr, Mo and W) have stronger adhesions than 
the other elements.
For any metal, the H-termination hinders the metal from
forming the electrodes because of weak bonding (see \S\ref{subsec:bonding_nature}).
In contrast, the O-termination makes quite different effects on the adhesion.
It weakens the adhesion for the noble metals (Au, Ag, Pt, and Pd), 
while it significantly strengthens for some transition metals (group-4,5,6)
by enhancing the ionic natures of bonding (see \S\ref{subsec:bonding_nature}).
O-terminated Cr and Ti are the promising candidates 
to achieve both higher DOS at the Fermi level, $D(\varepsilon_F)$, and anti-peeling strength.
It is remarkable that $W_\mathrm{sep}$ for O-terminated Cr is
around twice larger than that for no-terminated Ti.
  Comparing $W_\mathrm{sep}$ between no-termination
  and O-termination, it is interesting to note that
  $W_{\mathrm{sep}}$ increases from group 4 to 6
  for no-termination but such a clear tendency
  does not appear for O-termination.
  We revealed instead $W_{\mathrm{sep}}$ has
  a negative correlation with adhesion length
  clearly (Fig.~\ref{fig.length_wsep}).

\vspace{2mm}
Table \ref{tab.w_sep} compares our numerical results of $W_\mathrm{sep}$
and $d_\mathrm{adh}$ with those by previous studies 
\cite{2010GUO,2014MONACHON,2004QI}
for Ni, Cu, Al, and Ti. Note that our metal/diamond interface models
consider only monolayers, while the previous ones do multilayers.
Despite such a significant difference,
we found that our numerical results for Cu and Ni agree well
with the previous ones (their differences in $W_\mathrm{sep}$
and $d_\mathrm{adh}$ lie within $\sim 0.6$ J/m$^2$ and $\sim 0.04$ \AA~{},
respectively). On the other hand, our work overestimated $d_\mathrm{adh}$
values for Al and Ti. Interestingly, the overestimation affects
$W_\mathrm{sep}$ in the opposite manner, {\it i.e.}, $W_\mathrm{sep}$
increases (decreases) for Al (Ti).
This may be explained by the different nature of the bonding,
{\it i.e.}, covalent for Al and ionic for Ti as discussed in \S~\ref{subsec:bonding_nature}:
Covalent bonds generally tends to have longer bond lengths 
({\it e.g.}, sparse structure in diamond structure) 
than ionic ones
({\it e.g.}, NaCl structure).
The optimal adhesion length for Al would be
longer than the previous ones and closer to ours,
leading to the stronger adhesion.
  On the other hand,
  our predicted adhesion length for Ti seems to deviate
  from the optimal one, leading to the weaker adhesion.
  \begin{table*}
    \begin{center}
    \caption{
    Comparison of $W_\mathrm{sep}$ (J/m$^2$) between ours and
    literatures for several metals (Al, Cu, Ti, and Ni).
    The corresponding  $d_\mathrm{adh}$ values ($\AA$) are also
    given in parentheses.
    Note that the present metal/diamond interface models employ
    monolayer metallic sheets, while the previous ones did multi-layer
    mettalic sheets.
  }
  \begin{tabular}{lcccc}
      \hline
      & This work & H. Guo {\it et al.}\cite{2010GUO} & C. Monachon {\it et al.}\cite{2014MONACHON} & Y. Qi and L.G. Hector\cite{2004QI} \\
      \hline
      Al & 4.90~J/m$^2$~(2.09~\AA) & 4.08~J/m$^2$~(1.86~\AA) & N/A & 3.98~J/m$^2$~(1.86~\AA) \\
      Cu & 2.90~J/m$^2$~(2.06~\AA) & 3.36~J/m$^2$~(2.09~\AA) & 3.04~J/m$^2$~(N/A) & N/A \\
      Ti & 4.04~J/m$^2$~(2.18~\AA) & 5.77~J/m$^2$~(1.94~\AA) & N/A & N/A \\
      Ni & 5.65~J/m$^2$~(1.92~\AA) & N/A & 5.00~J/m$^2$~(1.96~\AA) & N/A \\
      \hline
    \end{tabular}
  \label{tab.w_sep}
  \end{center}
\end{table*}

\vspace{2mm}
We note that the present work did not 
take explicitly into account
the energy loss/gain by the lattice relaxation.
The metallic ion is located just above the
carbon atom in our slab model.
This corresponds to the situation where
the lattice of metallic sheet is forced to have
the same lattice constant as the diamond surface.
We may estimate the energy loss due to
this artificial distortion.
From the literature values of bulk modulus of metals,
we can roughly estimate the energy loss
that arises when
the lattice of metallic sheets is distorted towards that of diamond surface.
Taking Birch-Murnaghan equation of state, \cite{1947BIR} we estimate the loss
being 0.377 J/m$^2$ for Au, 0.015 J/m$^2$ for Pd,
0.073 J/m$^2$ for Ti, 0.342 J/m$^2$ for V,
and 0.078 J/m$^2$ for Ta.
Such energy losses are negligibly small compared
with $W_\mathrm{sep}(x)$ for metal sheets that show good adhesion.

\section{Conclusion}
\label{sec.conc}
The predicted adhesion energies indicates that 
the transition metal sheets can be better candidates 
for the electrodes on the diamond (111) surfaces, 
realizing more stable adhesion than conventional 
noble metal electrodes. 
DOS analysis does indeed confirm that some of those are 
realizing Ohmic contacts. 
For the termination atoms of diamond surfaces, 
hydrogen is predicted to be unbound with metallic sheets
while oxygen assists with the realization of more stable adhesion.
The trend of the relative adhession length for colavent length
can be reasonably explained in terms of the electronegativity.
It is also observed that adhession strength has a negative correlation
with adhesion length.
We find that Cr with oxygen-termination achieves the largest adhesion
in terms of work of separation,
while Ti with no-termination realizes the largest carrier density
in terms of DOS at the Fermi level.

\section{Acknowledgments}
The computations in this work have been performed 
using the facilities of the Center for Information Science in JAIST. 
This work was supported by the Kenya National Council
for Science and Technology (NCST, Now National Commission
for Science, Technology and Innovation (NACOSTI))
Grant No. NCST/5/003/4th CALL/050,
and the Computational Materials Sciences Group, Department of Physics,
University of Eldoret, Kenya. 
Authors, IM, NWM, and GOA wish to acknowledge the Center for
High Performance
Computing (CHPC) in Cape Town South Africa for computational resources.
KH is grateful for financial support from a KAKENHI grant (15K21023, 17K17762),
a Grant-in-Aid for Scientific Research on Innovative Areas (16H06439),
PRESTO (JPMJPR16NA) and the ``Materials research by Information Integration''
Initiative (MI$^2$I) project of the Support Program for Starting Up Innovation
Hub from Japan Science and Technology Agency (JST).
R.M. is grateful for financial support from 
TOKUYAMA science foundation, MEXT-KAKENHI 
grants 17H05478, the support by FLAGSHIP2020, 
MEXT for the computational resources, 
Project Nos. hp170269 and hp170220 at K-computer.

\bibliography{references}
\end{document}